\documentclass{elsart}
\usepackage{amssymb,amsmath,multirow,epsfig,graphicx,color}
\usepackage{color}

\usepackage{soul}
\begin{document} 
\begin{frontmatter}
\title{Bethe-Salpeter bound-state structure in Minkowski space}
\author[IFT]{C. Gutierrez},
\author[ITA]{V. Gigante},
\author[ITA]{T. Frederico},
\author[ROMA]{G. Salm\`e},
\author[PISA]{M. Viviani} and
\author[IFT,ITA]{Lauro Tomio}
\footnote{Corresponding author: tomio@ift.unesp.br}
\address[IFT]{Instituto de F\'isica Te\'orica, Universidade Estadual Paulista, 01156-970 S\~ao Paulo, SP, Brazil}
\address[ITA]{Instituto Tecnol\'ogico de Aeron\'autica, DCTA,12.228-900 S\~ao Jos\'e dos Campos, SP, Brazil.}
\address[ROMA]{Istituto Nazionale di Fisica Nucleare, Sezione di Roma, P.le A. Moro 2, 00185 Roma, Italy}
\address[PISA]{Istituto Nazionale di Fisica Nucleare, Sezione di Pisa, Largo Pontecorvo 3, 56100 Pisa, Italy}

\date{\today}
\maketitle
\begin{abstract}{ 
The quantitative investigation of  the scalar Bethe-Salpeter equation in Minkowski space, within the 
ladder-approximation framework, is extended to include the excited states. 
This study has been carried out for an interacting system composed by  two massive bosons exchanging a  
massive scalar, by adopting (i) the Nakanishi integral representation of the Bethe-Salpeter 
amplitude, and (ii) the formally exact  projection onto the  null plane.   Our  analysis,
on one hand, confirms  the reliability of the method already applied to the ground state and, 
on the other one, extends the investigation from the valence distribution in momentum space 
to the corresponding quantity in the impact-parameter space,
pointing out some relevant features, like (i) the equivalence 
between  Minkowski and Euclidean transverse-momentum amplitudes, and (ii)   
the leading exponential fall-off of the valence wave function in the impact-parameter space}.
\end{abstract}
\end{frontmatter}

\section{Introduction}

In the last decade and a half,  a quite effective tool is emerging for solving the 
Bethe-Salpeter equation (BSE)\cite{SB_PR84_51} directly in  Minkowski space, i.e.
avoiding to look for solutions in the Euclidean space by exploiting the Wick rotation
\cite{WICK_54}. The novel approach is based on the Nakanishi integral representation 
(NIR) of the $n$-leg transition amplitudes, that was proposed long time 
ago~\cite{nak63,nak69,nak71}. In particular, the Bethe-Salpeter (BS) amplitude can be formally written 
like the NIR for the $3$-leg amplitude, namely a proper folding of an unknown Nakanishi
weight function and a denominator that contains the analytic
structure \cite{KusPRD95,KusPRD97,SauPRD03,carbonell1,carbonell2,carbonell3,FSV1,FSV2}. To 
be practical, let us quickly mention the main features of the NIR for the BS amplitude: (i) 
the weight function is real and smooth for bound states 
(see \cite{FSV3} for the zero-energy scattering states), and  (ii) it depends upon
real variables, of which one is non-compact  and the others  are compact; (iii)  the
denominator must depend only upon the independent scalars that can be constructed
from the external momenta. Assuming the validity of the NIR for the bound-state case,
and taking advantage of the above mentioned features, one can exactly project onto the 
null-plane (see, e.g. \cite{SalPRC00,FreFBS00}) the BS
amplitude  integrating over the Light-front (LF) variable $k^-=k^0-k^3$  ($k^{+}= k^{0}+ k^{3}$ 
and ${\bf k}_{\perp}\equiv \{k^1,k^2 \} $) and formally obtain the so-called 
LF {\em valence wave function} (cf. \cite{CK_rev,Brodrev,FSV1}), i.e. the amplitude of the component
with the lowest number of constituents when the LF Fock expansion of the
interacting-system state is considered.  Remarkably, within the NIR approach, the LF 
valence wave function is given by a non-singular integral involving the Nakanishi 
weight function. This suggests to integrate on $k^-$  both sides of the BSE, 
getting an integral equation for the Nakanishi weight function.
If there exist solutions for this integral equation (this validates a posteriori the
previous assumption), then the BS amplitudes of bound states can be reconstructed. 
In particular, when the above procedure is applied to the BSE with the irreducible kernel in 
ladder approximation, a generalized eigen-equation for the Nakanishi weight function is obtained 
(see, e.g., Refs. \cite{carbonell1,FSV2} for the LF case), while for
the cross-ladder case one has to deal with a non-linear eigen-problem (see Ref. \cite{carbonell2}).
Notice that in Refs. \cite{KusPRD95,KusPRD97} solutions of the scalar BSE in ladder approximation 
have been obtained by using (i) standard variables (and not the LF ones), and (ii)
exploiting the uniqueness of the Nakanishi weight function.

Aim of  the present work is to carefully study both spectrum and 3D structure of the bound states, 
obtained by solving the ladder BSE for a system composed by two massive scalars interacting through
a massive scalar. Such an investigation is a natural extension of the  previous analysis of only 
the ground state \cite{FSV2}. In particular, the structure is studied by means of the 3D 
representation of the LF valence component, both in momentum and impact-parameter (IP) spaces. 
One of the motivations for starting a detailed analysis of the non-perturbative features of an interacting
system in momentum and IP spaces (see, e.g.\cite{BurIJMP03} for an introduction) is given 
by the increasing interest on this topic in hadronic physics, where  the valence component plays an 
important role in determining the dynamical properties of hadrons.  For instance, the  valence 
component is an important dynamical ingredient for evaluating  parton transverse-momentum distributions, 
which depend upon both the Bjorken momentum fraction $x$ and the transverse components of parton 
momentum~\cite{DiePRep03,BarPRep02}, or  parton density distributions in IP 
space, that can be related to the generalized parton distributions (see, e.g., Ref.~\cite{DiePRep03}).

Our paper is organized as follows. 
In Sec. \ref{MSBSE}, we quickly introduce  the general formalism (see, e.g.,
Refs. \cite{FSV1,FSV2,FSV3} for more details) and we present
a comparison between Minkowski and Euclidean results for the eigenvalues of the relevant
integral equation. In Sec. \ref{LFWF}, the
valence LF wave function and the corresponding density  distributions, 
evaluated  both in transverse-momentum space
and  impact-parameter one, are discussed, showing  our numerical results 
 for the available spectrum together with some interesting formal outcomes of our analysis.
 In Sec. \ref{END}, conclusions are drawn and some perspectives presented.

\section{Minkowski space solutions of the Bethe-Salpeter equation} 
\label{MSBSE}
 Let us recall the general formalism we have adopted to solve the BSE in Minkowski space.
As it is well known the BSE in momentum space for 
a relativistic bound state is given by the following homogeneous integral equation
\begin{equation}\label{bse}
\Phi(k,p)= G_{0}^{(12)}(k,p) \int \frac{d^4 k'}{(2\pi)^4} {\rm i} 
K(k,k';p)\Phi(k',p) \, ,
\end{equation}
where ${\rm i} \,K(k,k';p)$ is the interaction kernel that contains all two-body irreducible diagrams,
$p^\mu$ is the total momentum with the bound state mass given by $M=\sqrt{p^2}$.
In the present approach we do not consider the self-energy contribution,  so that 
$G_{0}^{(12)}(k,p)$ is the product of two free propagators,
\begin{equation}\label{free-prop}
G_{0}^{(12)}(k,p)=\frac{\rm i}{\left[(p/2+k)^2 -m^2 +{\rm i}\epsilon\right]} \frac{{\rm i}}{\left[(p/2-k)^2 -m^2 +{\rm i}\epsilon\right]}
,\end{equation}
with $m$ the constituent mass.
The BS amplitude for an $s-$wave state solution of Eq. (\ref{bse}) 
can be written in  terms of NIR as \cite{carbonell1,FSV1,FSV2}
\begin{equation}\label{NIR-bs}
\Phi(k,p)=-{\rm i} \int_{-1}^{1} dz' \int_{0}^{\infty} d \gamma' \frac{g(\gamma',z';\kappa^2)}
{[\gamma'+\kappa^2-k^2- p \cdot kz'-{\rm i}\epsilon]^3}, 
\end{equation}
where $\kappa^2\equiv m^2-M^2/4$.
By substituting (\ref{NIR-bs}) into (\ref{bse}) and integrating over $k^{-}$ on both sides,  one can
obtain the following generalized integral equation for the Nakanishi weight function 
(for details see Refs.~\cite{carbonell1,FSV1,FSV2}):
{\small
\begin{equation}\label{nakie}
\hspace{-0.2cm}\int_0^{\infty}\hspace{-0.2cm}d\gamma'
\left\{\hspace{-0.1cm}
\frac{g( \gamma', z; \kappa^2)}
{[ \gamma' + \gamma + z^2 m^2 + \left( 1 - z^2 \right) \kappa^2]^{2} }-
\hspace{-0.2cm}
\int_{-1}^{1}\hspace{-0.2cm}dz' V^{LF}(z,z',\gamma, \gamma') g(\gamma',z';\kappa^2)\right\}
=0
,\end{equation}\hspace{-0.2cm}
}
 where {\small
\begin{equation} \label{val1}
\int_0^{\infty}d\gamma'~\frac{g(\gamma',z;\kappa^2)}
{[\gamma'+\gamma +z^2 m^2 + \left( 1 - z^2 \right) \kappa^2]^2}=p^+
\int {dk^- \over 2 \pi} \Phi(k,p)=
{\frac{\sqrt{2}\, \psi(\xi,{\bf k}_\perp)} {\xi(1-\xi)}}, 
\end{equation}
}
with $\gamma=|{\bf k}_\perp|^2$, $\xi=(1 -z)/2$ and $\psi(\xi,{\bf k}_\perp)$ is the {\em valence 
light-front wave function} (the factor $\sqrt{2}$ comes from the symmetry of the problem; see, for example, 
\cite{FSV1}). In Eq. (\ref{nakie})
$V^{LF}$ is the Nakanishi kernel given in terms of the BS 4D kernel, by
{\small \begin{equation}
V^{LF}(z,z',\gamma, \gamma')\equiv {\rm i} p^{+} \int_{-\infty}^{\infty} \frac{dk^{-}}{2 \pi} G_{0}^{(12)}(k,p) \int \frac{d^4 k'}{(2 \pi)^4} 
\frac{{\rm i}K(k,k';p)}{[k'^2 + p\cdot k'z'-\gamma'-\kappa^2 + {\rm i}\epsilon ]^3}
.\end{equation}}
In this work we adopt the ladder approximation for the  BS kernel:
 \begin{equation}
\label{ladder-ker}
{\rm i}\,K^{(Ld)}(k,k')=\frac{{\rm i}\,(-{\rm i} \,g)^2}{ (k-k')^2-\mu^2+{\rm i}\,\epsilon} =
-{\rm i}~\frac{\alpha~(16 \pi m^2 )}{ (k-k')^2-\mu^2+{\rm i}\,\epsilon}
\, ,
\end{equation}
where $\alpha=g^2/(16 \pi m^2)$ and $\mu$ is the exchanged-scalar mass.
According to \cite{FSV2}, we have solved  Eq. (\ref{nakie}) by using a basis function 
expansion of the Nakanishi weight function, composed by  Laguerre polynomials 
$L_j(a\gamma)$ (with $j=0, 1, N_g$) for describing the $\gamma$-dependence (where 
$a$ is an appropriate parameter, as discussed in \cite{FSV2}) and  even 
Gegenbauer polynomials $C^{(5/2)}_{2\ell}(z)$ for the $z$ one (with $2\ell=0,2,...,2 N_z$). 
More specifically, for the $\gamma$-dependence we use an expansion in terms of the
functions ${\cal L}_j(\gamma) \equiv \sqrt{a} L_j(a\gamma) e^{-a\gamma/2}$, where
$\int_0^\infty d\gamma {\cal L}_i(\gamma){\cal L}_j(\gamma) = \delta_{ij}$.
The expansion in Gegenbauer polynomials is given in terms of the functions
$G_\ell(z)\equiv 3\sqrt{(2\ell)!\left(2\ell+\frac52\right)/\Gamma(2\ell+5)}(1-z^2)C_{2\ell}^{5/2}(z)$,
where $\int_{-1}^{1}dzG_\ell(z)G_{\ell'}(z)=\delta_{\ell\ell'}$. 
This last choice is dictated by the
symmetry property of the Nakanishi weight function 
$g(\gamma,z;\kappa^2)=g(\gamma,-z;\kappa^2)$, that is requested by the bosonic nature of the
adopted constituents. It should be recalled that a definite statistical property of the BS amplitude
avoids the so-called abnormal solutions of BSE, namely the ones 
 with negative norm \cite{nak63,nak69,KusPRD97},  that are associated with 
excitations in relative time of the bound states (see Refs.~\cite{alkofer,desplanques} for a
more recent discussion of the issue).  
Finally, the $z^2$ dependence of $g(\gamma,z;\kappa^2)$ entails a symmetry of
 the valence wave function, namely $\psi(\xi,{\bf k}_\perp)=\psi(1-\xi,{\bf k}_\perp)$.

 In our numerical approach, accurate convergence was achieved for the
ground state by using 14 Laguerre ($N_g=13$) and 10 Gegenbauer ($N_z=9$)
polynomials. For the excited states, the convergence was reached with
26 Laguerre and 10 Gegenbauer polynomials.
After introducing the basis function expansion and the ladder
approximation Eq. (\ref{ladder-ker}),  Eq. (\ref{nakie}) 
turns into the   matrix form of a generalized eigenvalue problem. 
In particular, one can symbolically write 
\begin{equation}\label{symbnak}
{\mathcal B}(M)\,g\, =\,\alpha {\mathcal A}(M)\,g ,
\end{equation}
where $g$ is the eigenvector.  Differently from the familiar non-relativistic case, in the 
eigen-equation (\ref{symbnak}) the role of eigenvalue is played by
the coupling constant $\alpha$, while
the mass of the system $M$ is a parameter that can be assigned, after fixing the 
exchanged-scalar mass $\mu$. In the standard way
of analyzing the BSE within the NIR framework \cite{KusPRD95,KusPRD97,SauPRD03,carbonell1,carbonell2,carbonell3,FSV1,FSV2}, 
one introduces the binding energy as  
\begin{equation}
B = 2m -M,
\end{equation}
which constrains the range of $B/m$ to the interval between $0$ and $2$, i.e. $0\le M/m\le 2$,
avoiding in this way the well-known instability of the $\phi^3$ model (see Ref.
\cite{gbaym}). In our NIR studies of the ladder BSE, Eq. (\ref{symbnak}) has a pivotal role.
First, after fixing $m$, $\mu$ and $B_{gr}$, it yields the coupling constant of the ground 
state, i.e. the smallest value of the coupling constant that we call $\alpha_{gr}$; secondly 
it allows one to calculate  the spectrum. Indeed, once we have found the
coupling constant $\alpha_{gr}$, one can find the excited state with respect to
$B_{gr}$ by slightly changing Eq. (\ref{symbnak}), as follows
\begin{equation}\label{symbnak1}
\lambda ~g= {1\over \alpha_{gr}}~{\mathcal B}(M)^{-1} {\mathcal A}(M)g
~~.\end{equation} 
In other words, after fixing $m$, $\mu$ and $\alpha_{gr}$, we search for  
values  $M=2m-B ~>~ M_{gr}$ that produce eigenvalues 
$\lambda=1$ (as trivially seen,   for $M=M_{gr}$ one has  $\lambda=1$).

\subsection{Comparing Minkowski and Euclidean eigenvalues}
In order to check the reliability of the computed masses for the excited states, 
we provide a comparison between the results of our calculations, obtained in the Minkowski 
space within the NIR, with those one can evaluate in the Euclidean space.
In Table \ref{comparison}, we show  the binding energies, in unit of the constituent mass $m$, for 
the  first, $B(1)/m$, and the second, $B(2)/m$, excited states, corresponding to a
ground state  $B(0)/m=1.9$ and different values of $\mu$. The choice of such a large binding energy is  
motivated by the fact that strongly-bound states should be affected by large relativistic effects. 
First, we have verified that the values of $\alpha_{gr}$ for the  binding 
energy of the ground state of $B(0)/m=1.9$, obtained with Euclidean- and Minkowski-space 
calculations are the same within our numerical accuracy. Then, we have computed the excited
state energies given in Table \ref{comparison}, achieving a very satisfactory agreement
between the results evaluated in the two spaces. As a remark on the numerical procedure, 
it should be pointed out that for values of $\mu/m$ smaller than 0.05 or  
$B/m\,< \,0.01$ the convergence is quite slow, and it is needed an extrapolation of the 
results with respect to $N_g$, in order to  accurately determine the eigenvalues.

\begin{figure}[h]
\begin{center}
\includegraphics[scale=0.5]{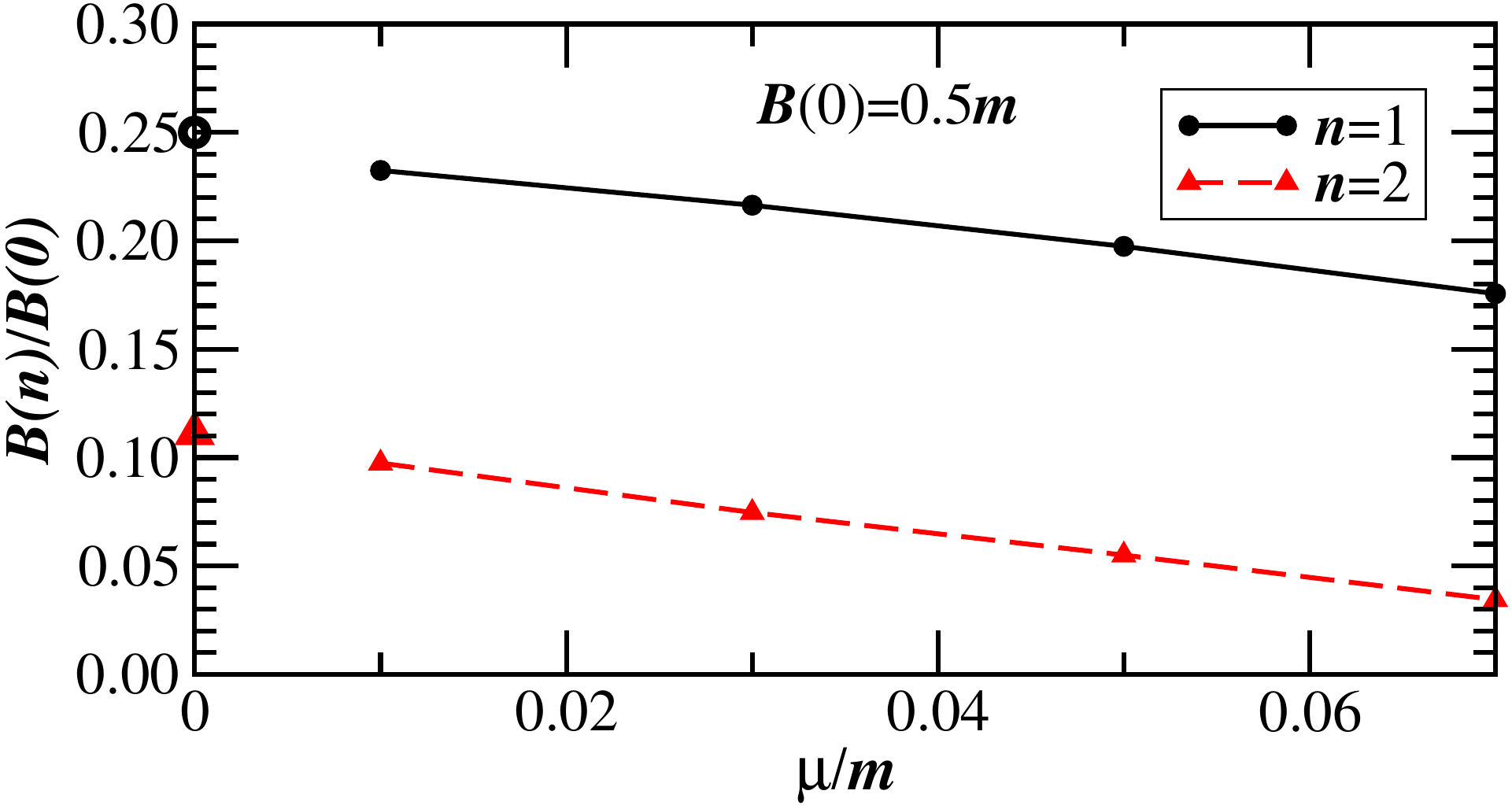}
\end{center}
\caption{
Energy ratios $B(n)/B(0)$ vs $\mu/m$ for the first (solid line with bullets) 
and second (dashed line with triangles) excited states. The  symbols on the lines are the
values obtained through Eq. (\ref{symbnak1}), while the circle ($n=$1) and  the triangle ($n=$2), 
at the origin, represent the corresponding non-relativistic limits, given by 
Eq.~(\ref{nrB}) with $ B_{nr}(0)/m=0.25\alpha^2 $.}
\label{ratio}
\end{figure}

It is also important to show the behavior of energy ratios, $B(n)/B(0)$ with $n\geq0$, for small $B/m$ 
and $\mu \rightarrow 0$. For a bound state composed for two spinless bosons exchanging a massless 
scalar boson, the corresponding relativistic expression 
in lower orders of $\alpha$, as derived in \cite{1973-Feldman}, is 
\begin{equation}
B(n)=\frac{m}{4} \frac{\alpha^2}{(n+1)^2}\left[1+\frac{4\alpha}{\pi}{\rm ln}\alpha \right]
+ ... \;\;\;  (n \geq 0) .
\label{nrB}
\end{equation}
The first term is the non-relativistic limit. 
As verified in Fig.~\ref{ratio}, { where $B(1)/B(0)$ and $B(2)/B(0)$
 are shown for small values of $\mu/m$, the energy ratios } are consistent 
with the non-relativistic limit. { Moreover, the agreement for $\mu\to0$ between
the relativistic and non-relativistic eigenvalues is observed only 
 for small values of $n$. Indeed,  as binding energies increase  relativistic
 effects become larger and larger}.

\begin{table}[h]
\begin{center}
\caption{ Comparison of the spectra obtained in the Euclidean space  in the Minkowski 
one, by varying $\mu/m$ and, consequently,  $\alpha_{gr}$, but taken fixed
 the value of  the  ground-state 
binding energy  to $B(0)/m=1.9$.}\label{comparison}
\begin{tabular}{cccc}
\hline\hline
 $(\mu/m , \alpha_{gr})$  &             & \textbf{Euclidean} & 
\textbf{Minkowski} \\ \hline \hline
\multicolumn{1}{c}{\vspace{-0.4cm} }	 & $B(1)/m$   & 0.258 & 0.259   \\
\multicolumn{1}{c}{\vspace{-0.4cm} (0.05, 6.324)}  &&& \\  
\multicolumn{1}{c}{ }  & $B(2)/m$   & 0.090 & 0.090   \\ \hline
\multicolumn{1}{c}{\vspace{-0.4cm}}          & $B(1)/m$   & 0.220 & 0.221   \\ 
\multicolumn{1}{c}{\vspace{-0.4cm} (0.1, 6.437)}  &&&   \\ 
\multicolumn{1}{c}{ } & $B(2)/m$   & 0.051 & 0.050   \\ \hline
\multicolumn{1}{c}{\vspace{0cm}} (0.5, 8.047)~~  & $B(1)/m$   & 0.0082& 0.0082  \\ 
\hline\hline
\end{tabular}
\end{center}
\end{table}

\section{ Valence light-front wave function and momentum distributions}
\label{LFWF}
It is attractive to perform numerical comparisons  of dynamical quantities
that in perspective could be useful for an experimental investigation of actual
interacting systems. In view of this, from   
the valence LF wave function introduced in Eq. (\ref{val1}) (see the next 
subsection for the numerical results),  one can define both the probability 
distribution to find a  
constituent with LF longitudinal fraction $\xi=p^+_i/P^+$,  given by
\begin{equation}
\varphi(\xi)= {1 \over 2 (2 \pi)^3}~  {1\over \xi(1-\xi)}~ \int 
 d^2{\bf k}_\perp~
 \Bigl [\psi(\xi,{\bf k}_\perp)\Bigr]^2 \, ,
  \label{phixi}\end{equation}
and the probability distribution to find a  
constituent with LF transverse momentum $k_\perp=|{\bf k}_\perp|$, that reads
 \begin{equation}
 {\mathcal P}(k_\perp)= {1 \over 4(2 \pi)^3}~  \int_0^1 {d\xi\over \xi(1-\xi)}~  
 \int_0^{2\pi} d\theta~
 \Bigl [\psi(\xi,{\bf k}_\perp)\Bigr]^2 \, .
 \label{probgam}
 \end{equation}
Both LF distributions are normalized to the probability of the valence component,
once the BS amplitude  itself 
is properly normalized (see Ref. \cite{lurie} for a general discussion and Ref. \cite{FSV2}
for the application within the NIR). Such a probability yields 
the probability to find the valence contribution in the LF Fock expansion of the interacting 
two-scalar state (see, e.g.,  \cite{Brodrev,FSV1,dae}).  As a matter of fact, one has 
\begin{equation}
 P_{val}= ~ \int_{0}^1 d\xi \, \varphi(\xi)\, = \int_{0}^{\infty} dk_\perp~ {\mathcal P}(k_\perp).
 \label{pval}\end{equation} 
Notice that ${\bf k}_\perp$ can be associated with the intrinsic transverse momentum,
in the frame where ${\bf p}_\perp=0$, which is allowed by the covariance of our 
description.

Although we have discussed the issue of the proper normalization of the valence state,
in what follows we are interested on the overall 3D structure of the valence wave function, and therefore we have
simply adopted  an arbitrary normalization. 

\subsection{Momentum space valence wave function for  excited states }

\begin{figure}[h]
\begin{center}
\includegraphics[scale=0.33]{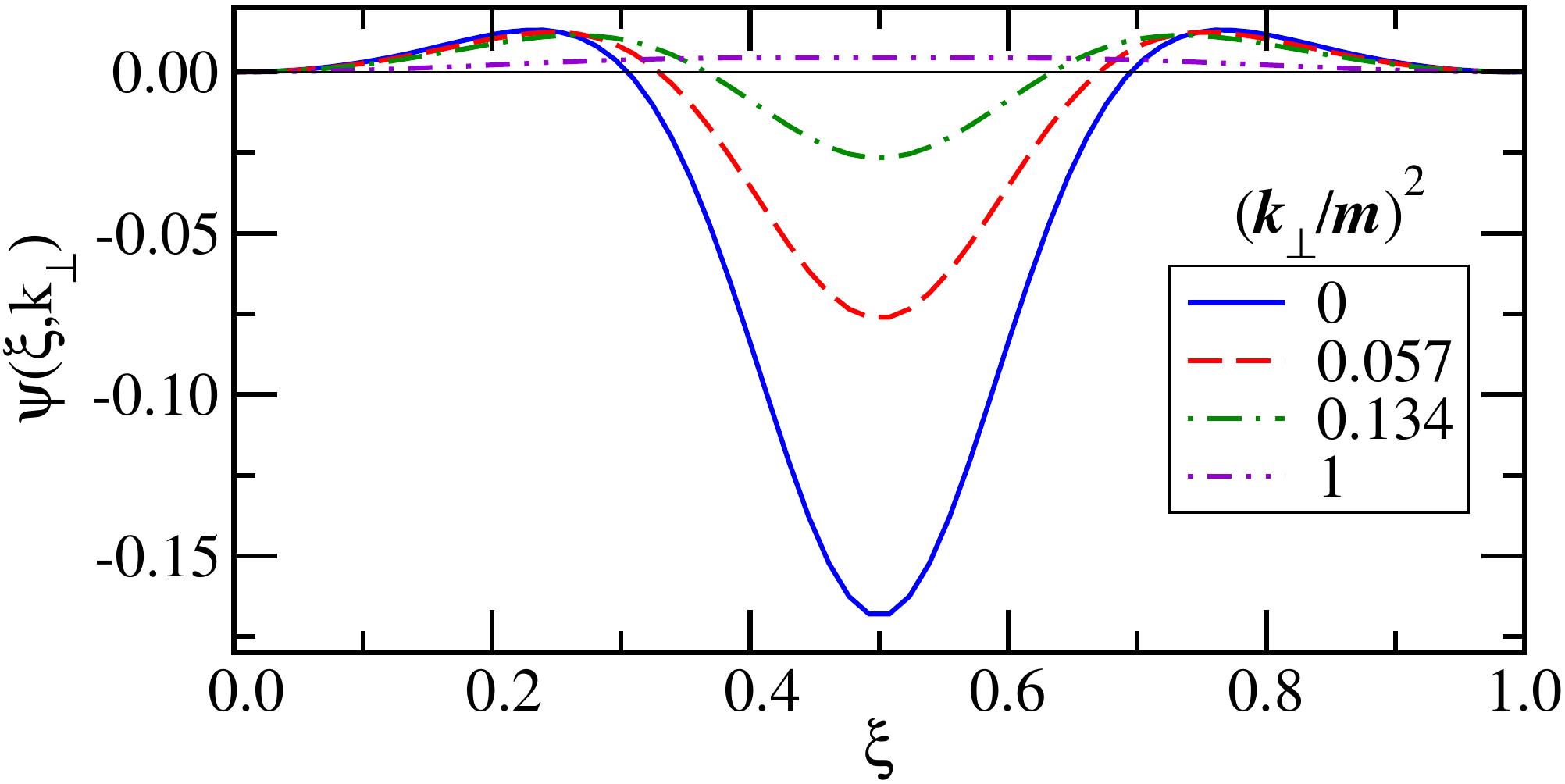}
\includegraphics[scale=0.33]{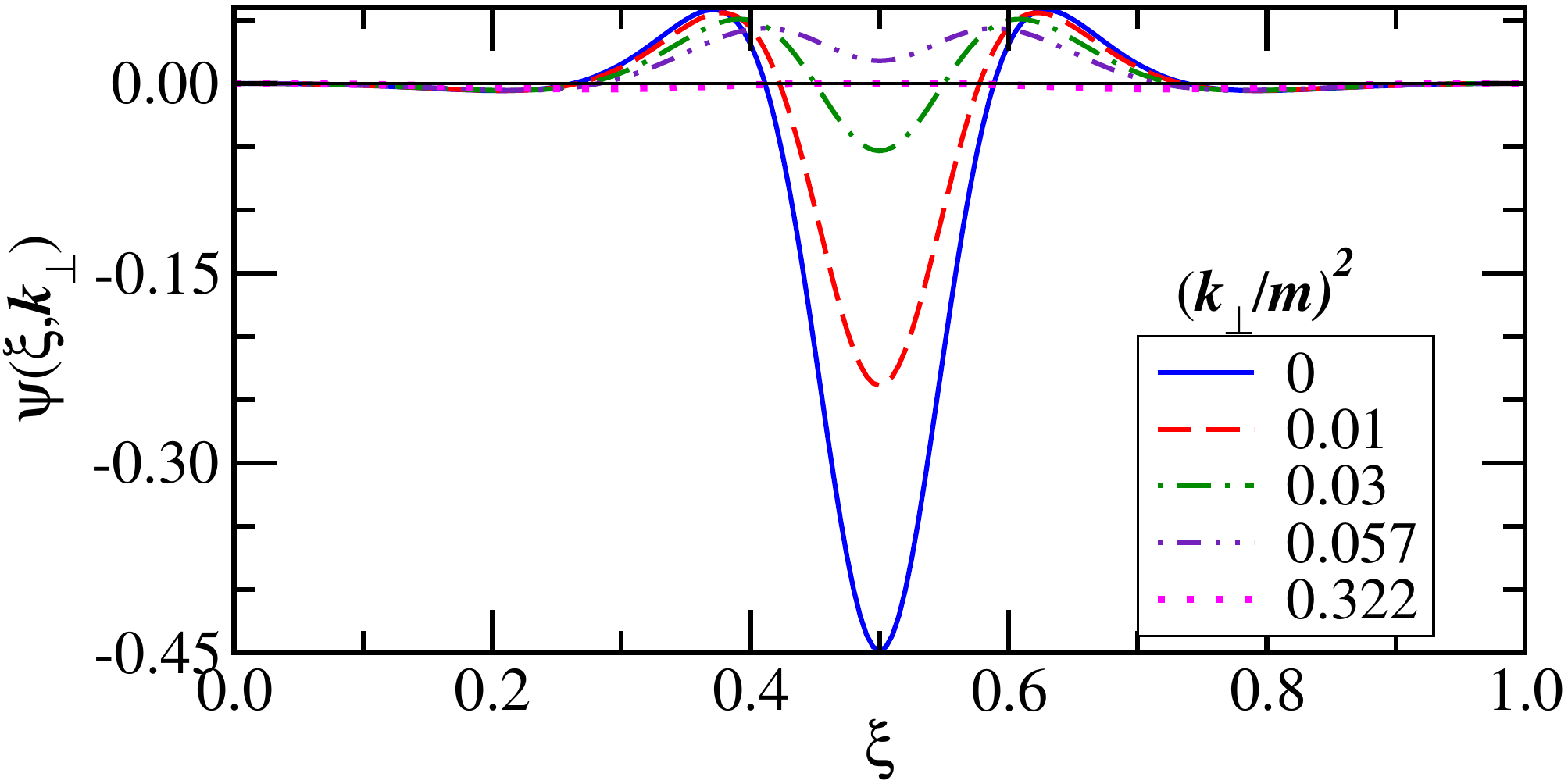}
\end{center}
\caption{
{The valence wave functions vs $\xi$ with fixed values of $(k_\perp/m)^2$, for the first (left panel) and second (right panel) excited states,
with $B(1)/m = 0.22$ and $B(2)/m = 0.05$, respectively, obtained from (\ref{symbnak1}) with  $\mu/m= 0.1$ and  $\alpha$=6.437. } 
}
\label{lf-g}
\end{figure}

\begin{figure}[h]
\begin{center}
\includegraphics[scale=0.33]{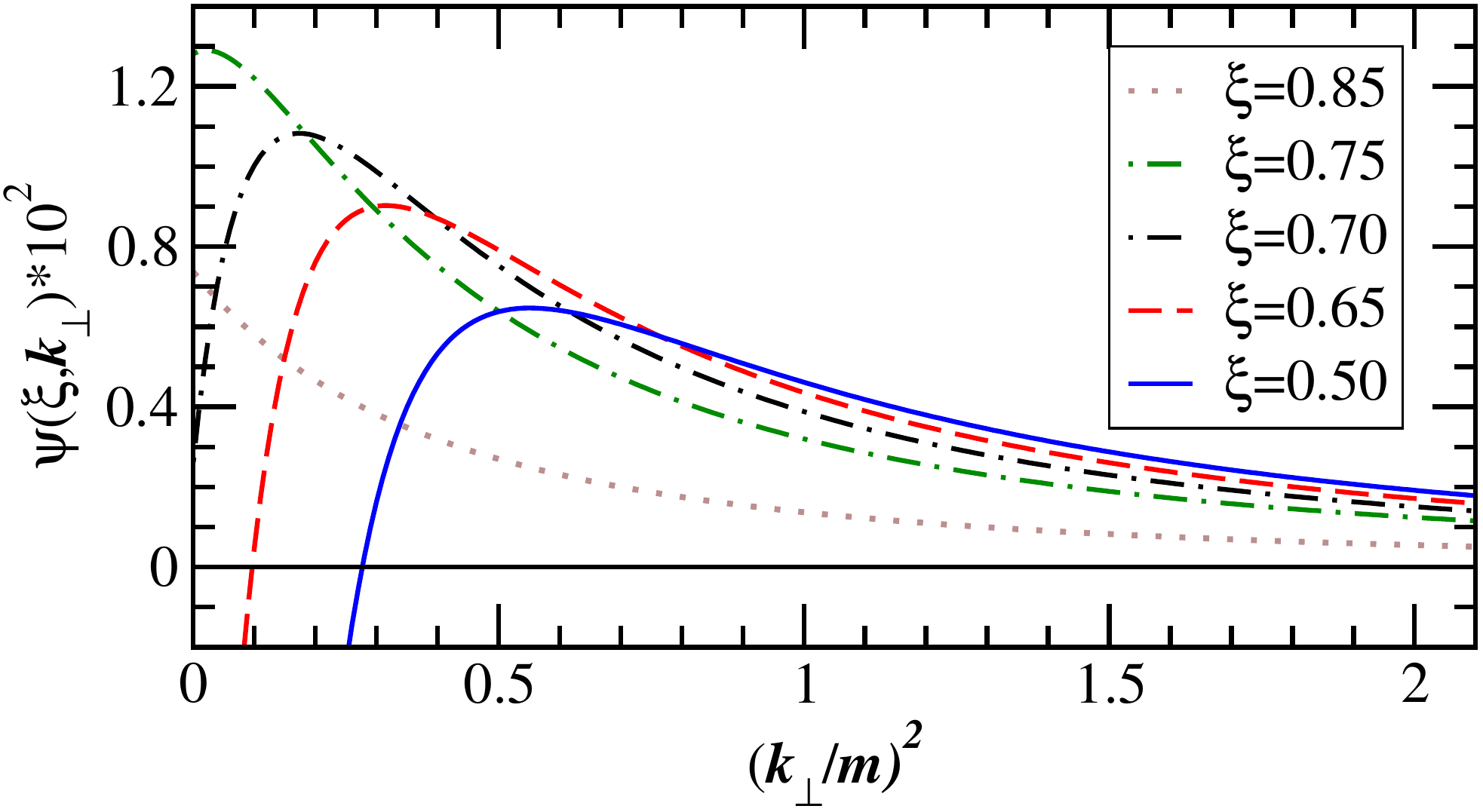}
\includegraphics[scale=0.33]{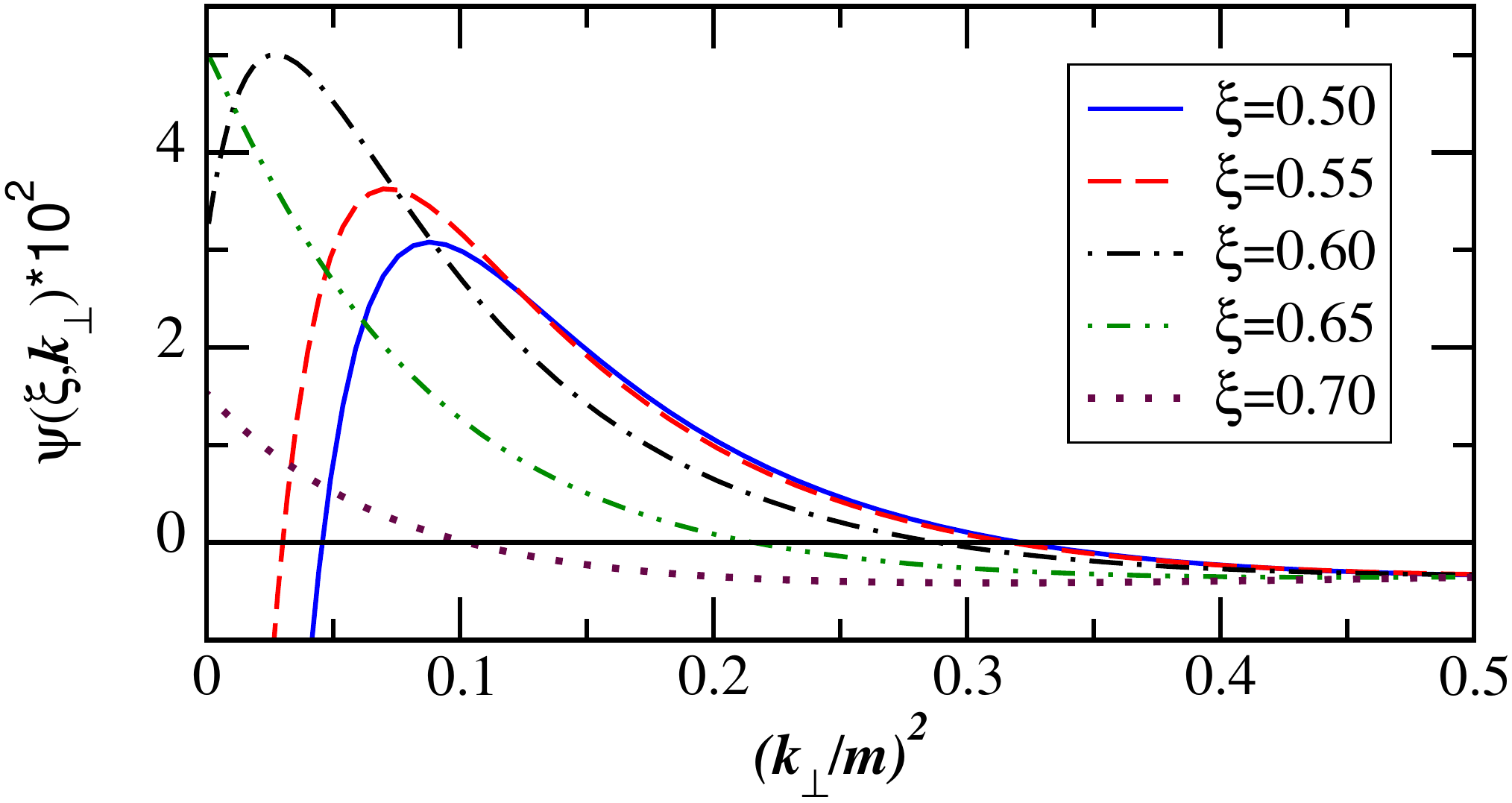}
\end{center}
\caption{
{The valence wave functions vs $(k_\perp/m)^2$ with fixed values of $\xi$, for the first (left panel) and second (right panel) excited states,
with $B(1)/m = 0.22$ and $B(2)/m = 0.05$, respectively, obtained from (\ref{symbnak1}) with  $\mu/m= 0.1$ and  $\alpha$=6.437. }
\label{lf-g1}
}
\end{figure}

{In Figs.~\ref{lf-g} and ~\ref{lf-g1}}, we present the LF wave function of the { first (left panels) 
and second (right panels)} excited states, corresponding
to the case $\mu/m$=0.1, $\alpha_{gr}$=6.437, $B(1)/m$=0.22 and $B(2)/m$=0.05
(see Table \ref{comparison}). As clearly shown, the wave function displays the typical 
feature of the first and second excited states, i.e. one and two \textit{nodes}, respectively. 

By a direct inspection of the corresponding panels for  the first excited state 
in Fig.~\ref{lf-g} { and for  the second  excited state 
in Fig.~\ref{lf-g1}}, one observes that, in the plane 
($\xi\, ,\, k_\perp/m$), the node structure is present 
for $(k_\perp/m)^2 < 1$ and  $\xi < 0.75 $, and  it is symmetric with respect to $\xi=1/2$.
In particular, the node structure moves toward $\xi=1/2$ 
 as $k_\perp$ increases. Such a behavior can be naively expected  when Cartesian three-momenta
  are adopted. As a matter of fact, the relation between Cartesian and LF components is
  \begin{equation}
{\bf k}^2=\frac{k^2_\perp+m^2}{4\xi(1-\xi)}-m^2 .
\label{cmom}
\end{equation}
 If  we assume   a dependence upon ${\bf k}^2$ for the excited-state valence wave function
 (instead of the actual dependence upon  
 $\xi$
 and $k_\perp$, separately), i.e. the same dependence  found
 in  phenomenological valence wave functions  widely
adopted for describing ground states (as the one exploited in  the discussion of the nucleon form factors in Ref.
\cite{Brodrev}), then the behaviors shown in 
{Figs. \ref{lf-g} and \ref{lf-g1}, for the node structures and asymptotic behaviors of the states,} 
become quite reasonable. 
Indeed, the assumed excited-state valence wave functions have to display   a  node at a fixed value for 
 ${\bf k}^2$, and therefore according to { Fig.~(\ref{lf-g})}, 
for increasing $k_\perp$ the variable $\xi$ is constrained to approach
$1/2$ (i.e. the maximal value of $\xi(1-\xi)$), in order to take (almost) constant 
${\bf k}^2$.  
In conclusion, the correlation between the LF components, 
$\xi$ and $k_\perp$, in determining the node position can be largely explained 
by the rotational invariance of the phenomenological wave functions,
 if they depend upon ${\bf k}^2$. Notably, our   calculations, genuinely in Minkowski space, actually confirm 
the overall expectation, based on a simple phenomenological Ansatz, that takes into account
the rotational invariance. It should be reminded that, within a LF framework, the rotational invariance  
can be fully recovered only if the whole Fock expansion is considered~\cite{Brodrev}.   We just add that
the second node  present in the right panel of Fig. \ref{lf-g} is hard to be seen given the scale of the plot.

\begin{figure}[h]
\begin{center}
\includegraphics[scale=0.33]{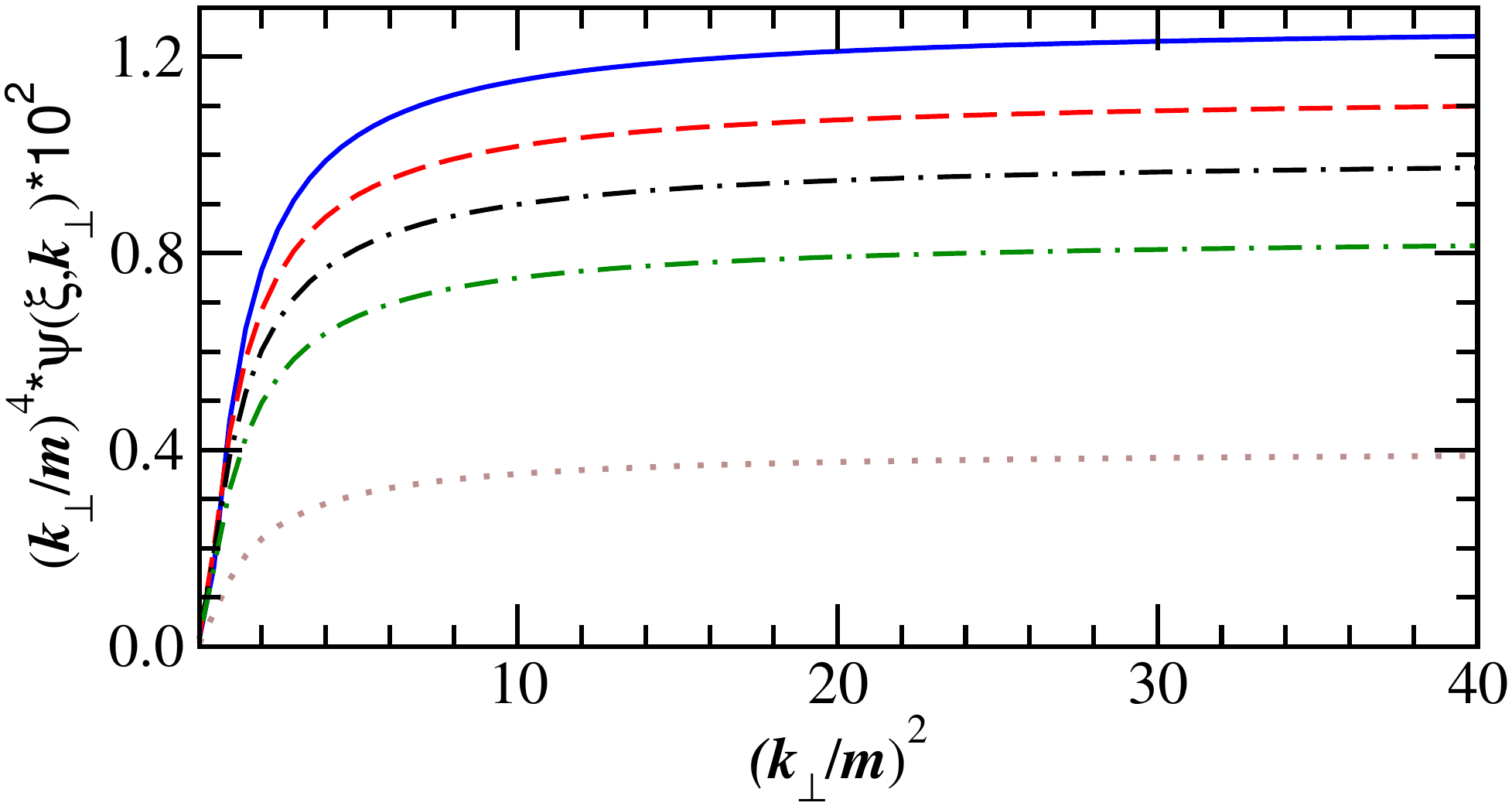}
\includegraphics[scale=0.33]{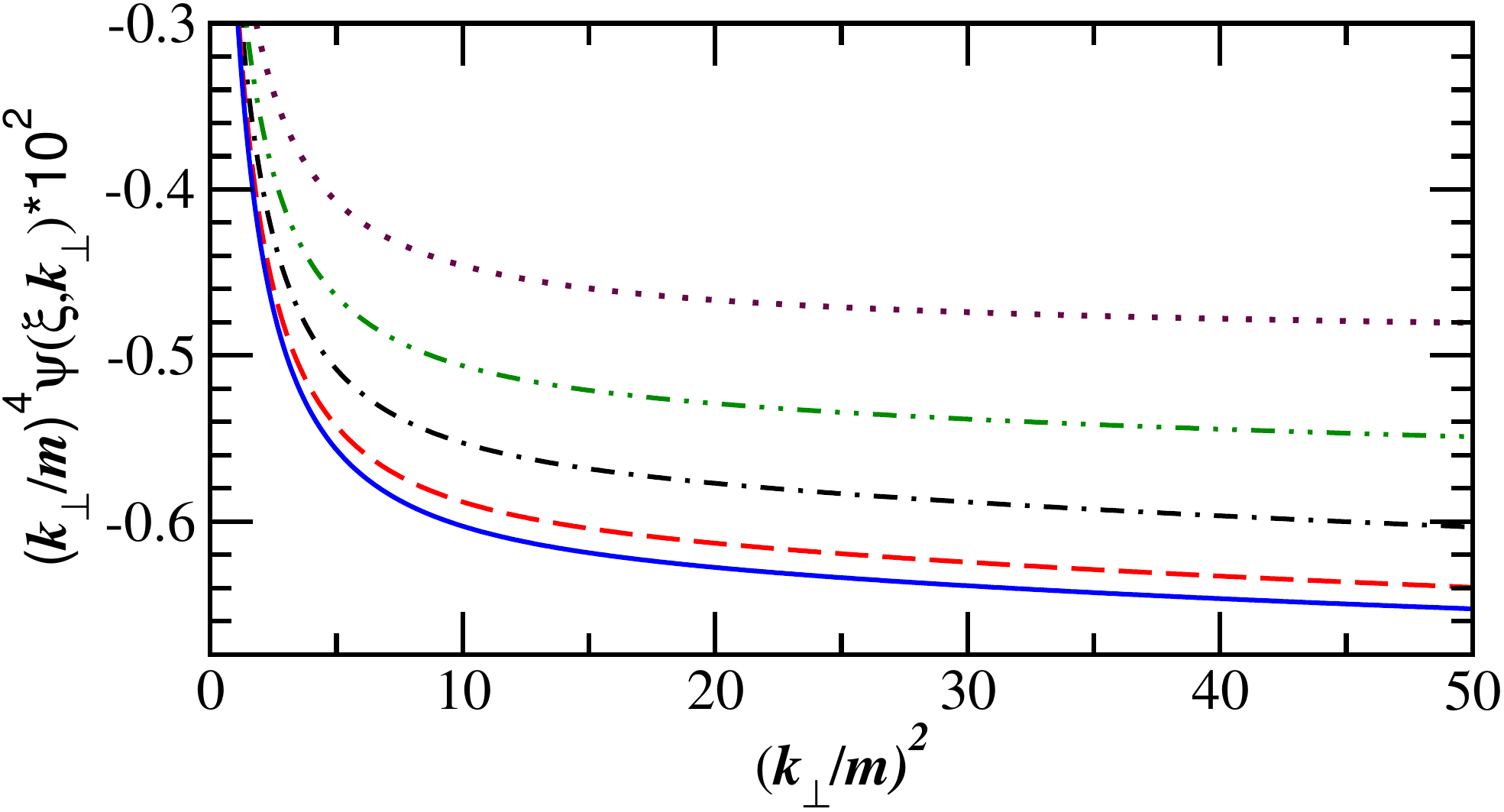}
\end{center}
\caption{
The asymptotic $k_\perp$ behaviors of the first (left frame) and second (right frame) excited states
are shown, using the same label convention as given in Fig.~\ref{lf-g1}. 
}
\label{lf-g2}
\end{figure}
From Eq. (\ref{val1}), one can obtain the asymptotic behaviors, $k_\perp\rightarrow \infty$, of LF valence wave function for 
bound states, which are given by
\begin{equation}\label{asymptot}
\psi(\xi,{\bf k}_\perp)\rightarrow {k_\perp^{-4}} \,{C(\xi)}\, .
\end{equation}
Such behavior is explicitly shown in Fig.~\ref{lf-g2}  for the first two excited states
(for ground-state, see Ref.~\cite{FSV2}).
 It should be emphasized  that independently of the value of $\xi$ the wave function is damped as $\sim k_\perp^4$.

We close  this subsection by discussing the equivalence of the transverse-momentum amplitudes in 
Minkowski and  Euclidean spaces~\cite{SalPRC00}, respectively defined as 
\begin{eqnarray}\label{tma}
\phi^T_{M}(\mathbf{k}_\perp)&\equiv&\int dk^0dk^3\Phi(k,p)= \frac12\int
dk^+dk^-\Phi(k,p)\;\;  {\rm and}\\
\phi^T_E(\mathbf{k}_\perp)&\equiv&{\rm i}\int dk^0_Edk^3\Phi_E(k_E,p),
\label{tea} \end{eqnarray}
where $\Phi_E(k_E,p)$ is obtained from $\Phi(k,p)$  after applying the Wick rotation with 
$k^0\to {\rm i} k^0_E$. 

Notably, within NIR, one can easily prove that
$\phi^T_{M}(\mathbf{k}_\perp)=\phi^T_E(\mathbf{k}_\perp)$, since one can exploit the explicit
expression of the  analytic dependence of the BS amplitude, as given in Eq. (\ref{NIR-bs}).
 As a matter of fact,  
choosing  the rest frame,  one can straightforwardly see that
the zeros of the Nakanishi denominator in the complex plane of $k_0$ are given by:
\begin{equation}
k_0=-\frac{M\,z'}{2}\pm\sqrt{\frac{M^2\,z'^2}{4}+\gamma' +\kappa^2+k_3^2+k_\perp^2-{\rm i} \epsilon} \, ,
\end{equation}
with $z'$  $\in[-1,1]$ and  $\gamma'$  $\in[0,\infty]$. Therefore,
 Eq. (\ref{NIR-bs}), as a function of complex $k_0$, has two cuts, 
 with branch-points  at
\begin{equation}
 k_{0\pm}^{\text{b}}=
\pm \left(\frac{M}{2}-\sqrt{\frac{M^2}{4}+\kappa^2}\right) \, .
\end{equation}
\begin{figure}[h]
\begin{center}
\includegraphics[scale=0.7]{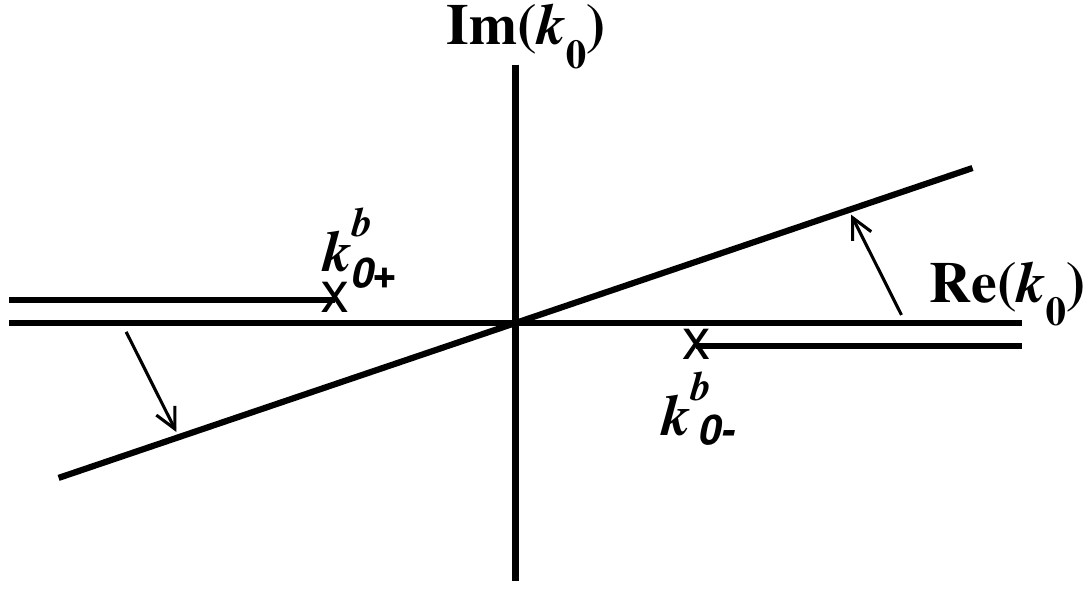}
\end{center}
\caption{
 Analytic structure of the BS amplitude in the complex plane of $k_0$, 
showing the left- and right-hand cuts with the corresponding branch points
$k_{0\pm}^{\text{b}}$. The rotation path of the $k_0$-integration contour is 
also shown for the transverse amplitude (\ref{tma}). }
\label{comprot}
\end{figure} 

Recalling that $\kappa^2$ is positive { for  bound states}, 
one can show that at the branch-point $ k_{0+}^{\text{b}}$ a cut 
starts in the upper half-plane for  
$\text{Re}\,k_0<0$, while at the branch-point  $ k_{0-}^{\text{b}}$ the cut
is placed in the lower   half-plane for
positive values of $\text{Re}\,k_0$, as shown in Fig. \ref{comprot}. 
If, in Eq. (\ref{tma}), where  the Minkowski space is adopted, one
considers the  integration variable $k_0$ as a complex one, i.e. $k_0=|k_0|
e^{i\theta}$, and rotates the angle $\theta$ 
up to  $90^o$, no singularities are crossed  (cf. Fig. \ref{comprot}).  
Furthermore, assuming that the BS amplitude drops out  fast enough  for 
large complex $|k_0|$, the Cauchy theorem holds and the
 Wick rotation \cite{WICK_54} can be applied for computing the transverse 
 amplitude. Namely, one can adopt a  new integration path, along a purely
 imaginary $k_0$, without dealing with any  singular integrals.  Consequently,
 the Minkowski and Euclidean transverse amplitudes, given 
 by Eqs.~(\ref{tma}) and (\ref{tea}) are formally equivalent.
 
\begin{figure}[h]
\begin{center}
\includegraphics[scale=0.33]{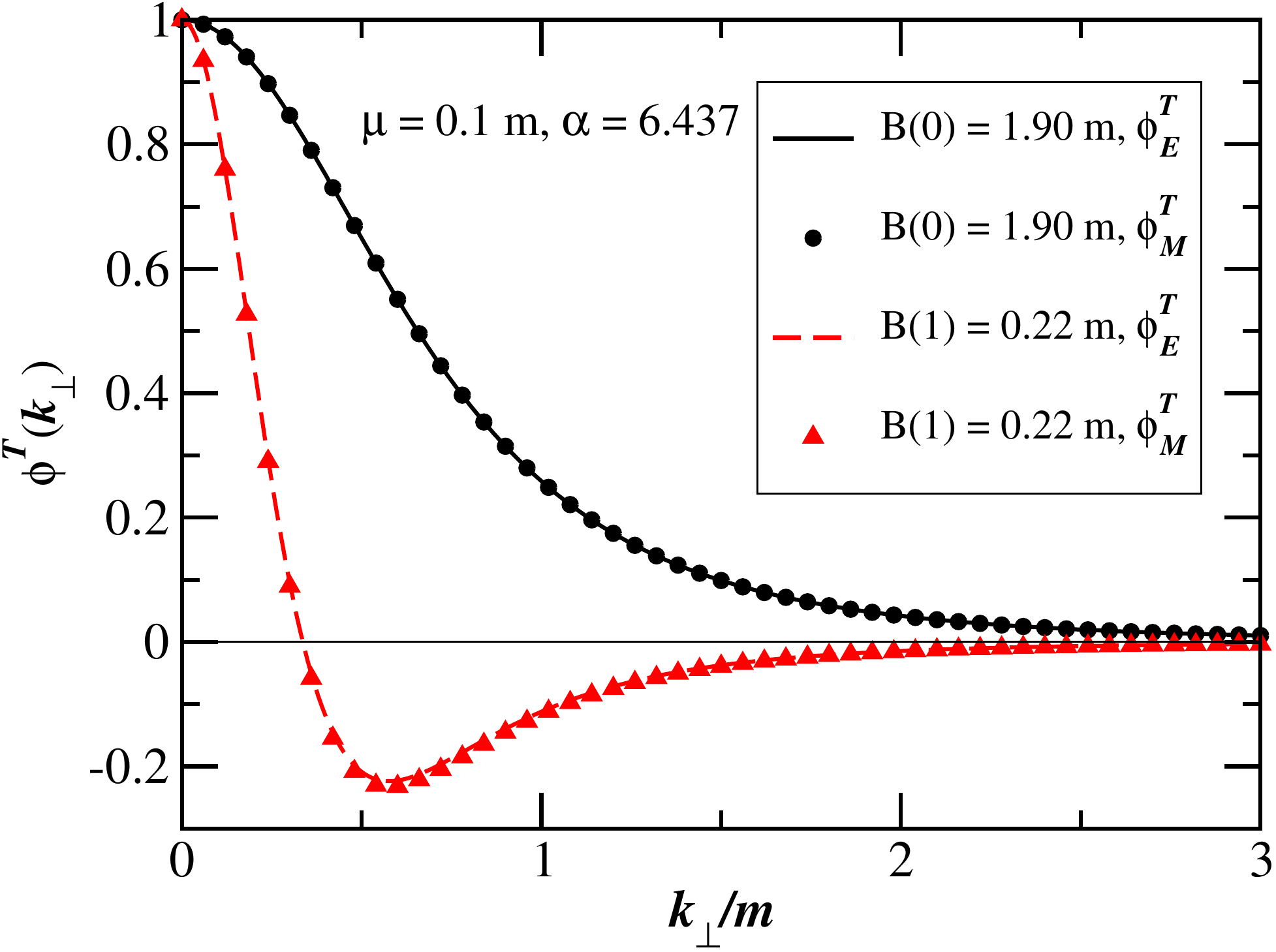}
\includegraphics[scale=0.33]{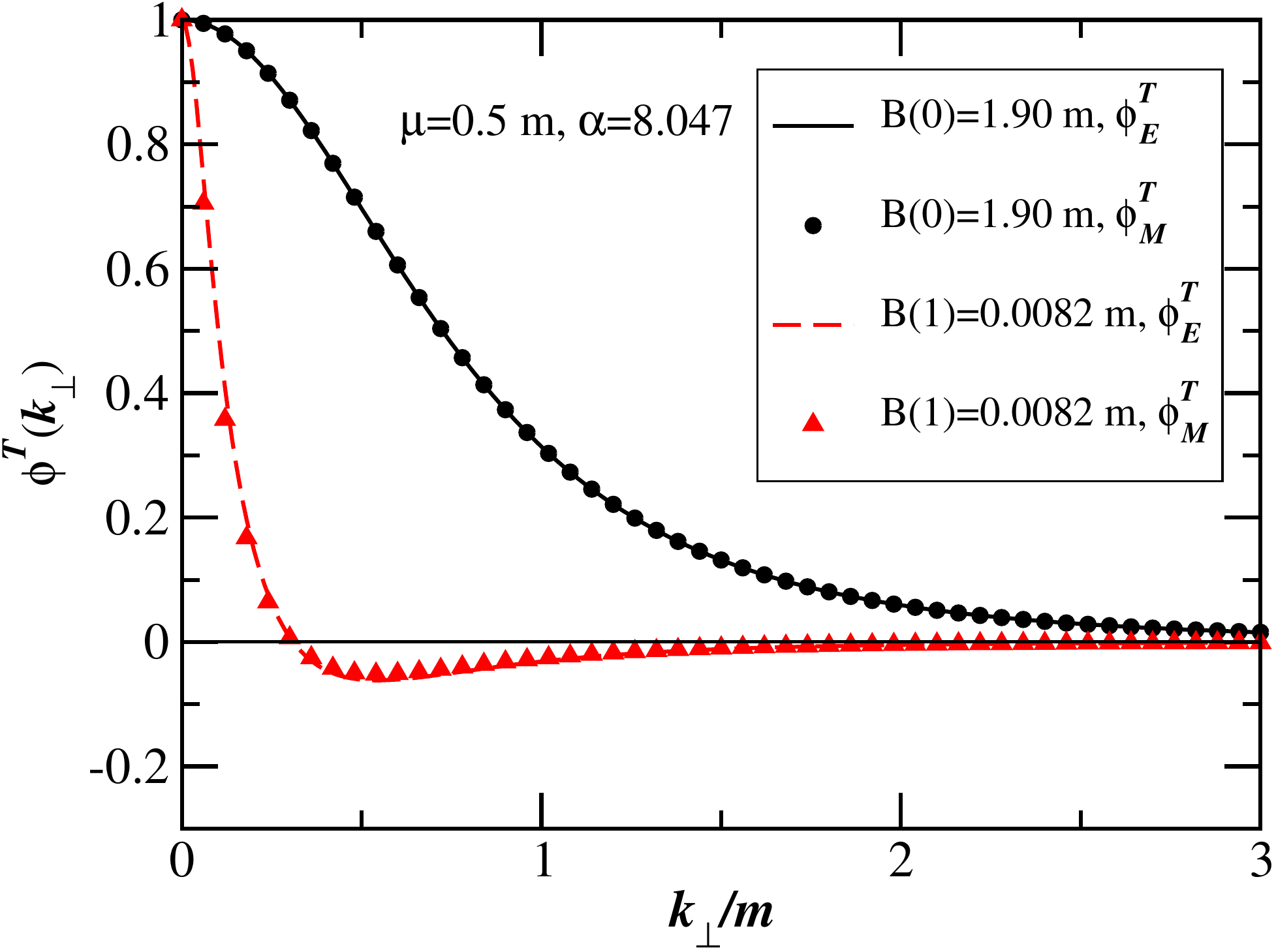}
\end{center}
\caption{Transverse momentum amplitudes $s-$wave states, 
in Euclidean and Minkowski  spaces, vs $k_\perp$, for both ground- 
and first-excited states, and two values of $\mu/m$ and $\alpha_{gr}$ (as indicated in the insets). The amplitudes 
$\phi_E^T$ and $\phi_M^T$, arbitrarily normalized to 1 at the origin, are not easily distinguishable.}
 \label{td}
\end{figure}

The quantitative  comparison for the cases $\mu/m$=0.1 and 0.5,
 with $\alpha_{gr}$ taken from Table \ref{comparison} 
(recall that one has always $B(0)/m=1.9$), is presented in Fig. \ref{td}, showing  
 a  very good agreement between   the transverse amplitudes, within the accuracy of our numerical
 approaches.  It is worth noticing that  such an equivalence gives an
  additional confidence in NIR, since it should be emphasized that the  Euclidean
solutions of BSE are not obtained by assuming  the  NIR for BS amplitudes. Therefore the
comparison in Fig. \ref{td} should be considered as a further check  
of the reliability of the NIR itself, at least at the ladder level, besides the passed tests 
for both eigenvalues \cite{KusPRD95,KusPRD97,carbonell1,FSV2} and  scattering 
lengths \cite{FSV3}.  
Moreover, Fig. \ref{td}  illustrates nice and general features of the transverse  amplitudes,
 that appear when    the binding energies change. As a
matter of fact, the position of the  node  in the first
 excited state moves toward smaller values of $k_\perp$ 
 as the binding energies decreases, i.e.
 from the left  panel 
($B(1)/m=0.22$)
to the right panel ($B(1)/m=0.0082$). Analogously, the amplitudes themselves 
decrease more 
quickly in momentum space. Both features can be explained by the increase
of   size of the 
bound state 
when the binding energy decreases.

\subsection{ Valence LF wave function in the impact-parameter space}

\begin{figure}[thb]
\begin{center}
\includegraphics[scale=.13]{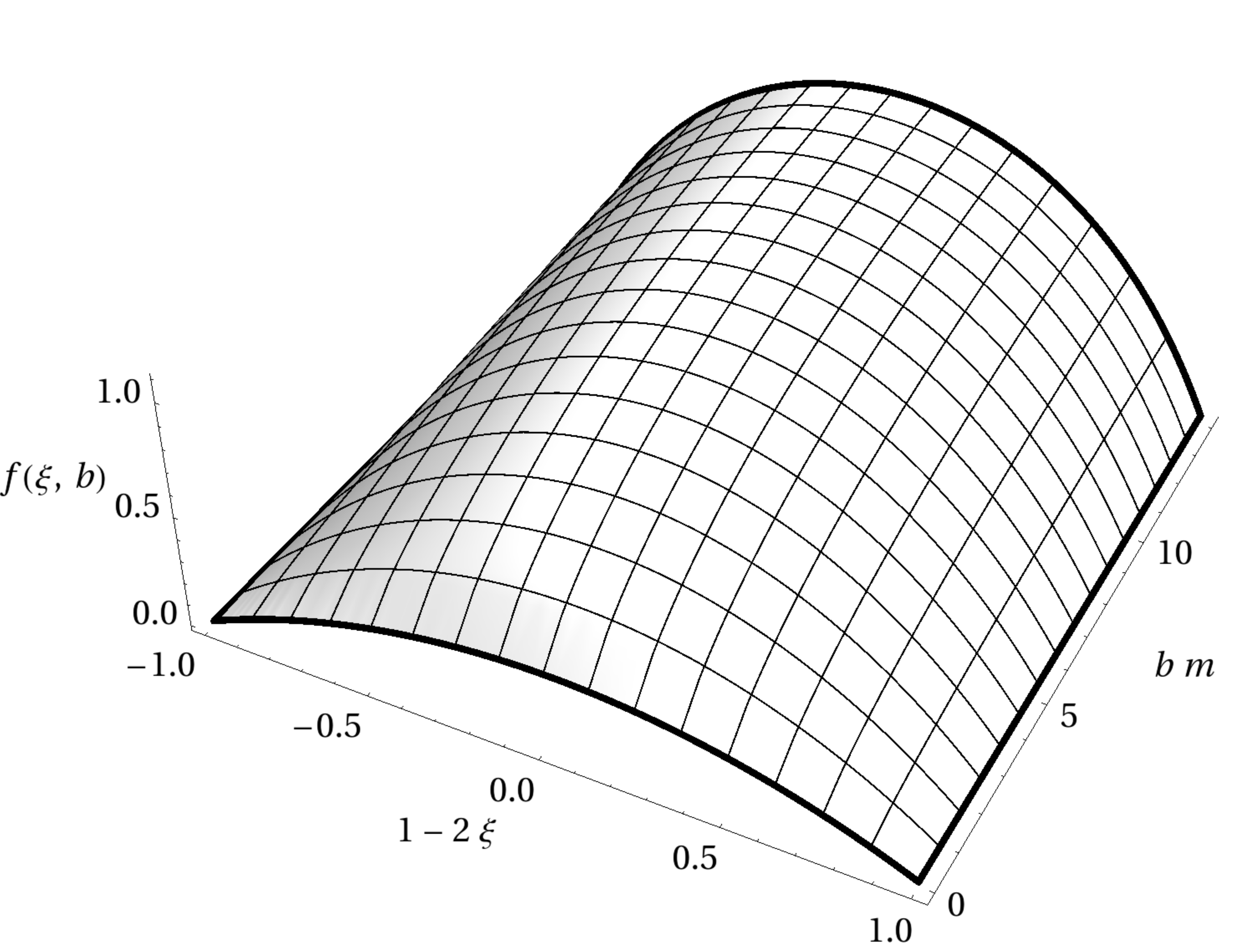}
\includegraphics[scale=.13]{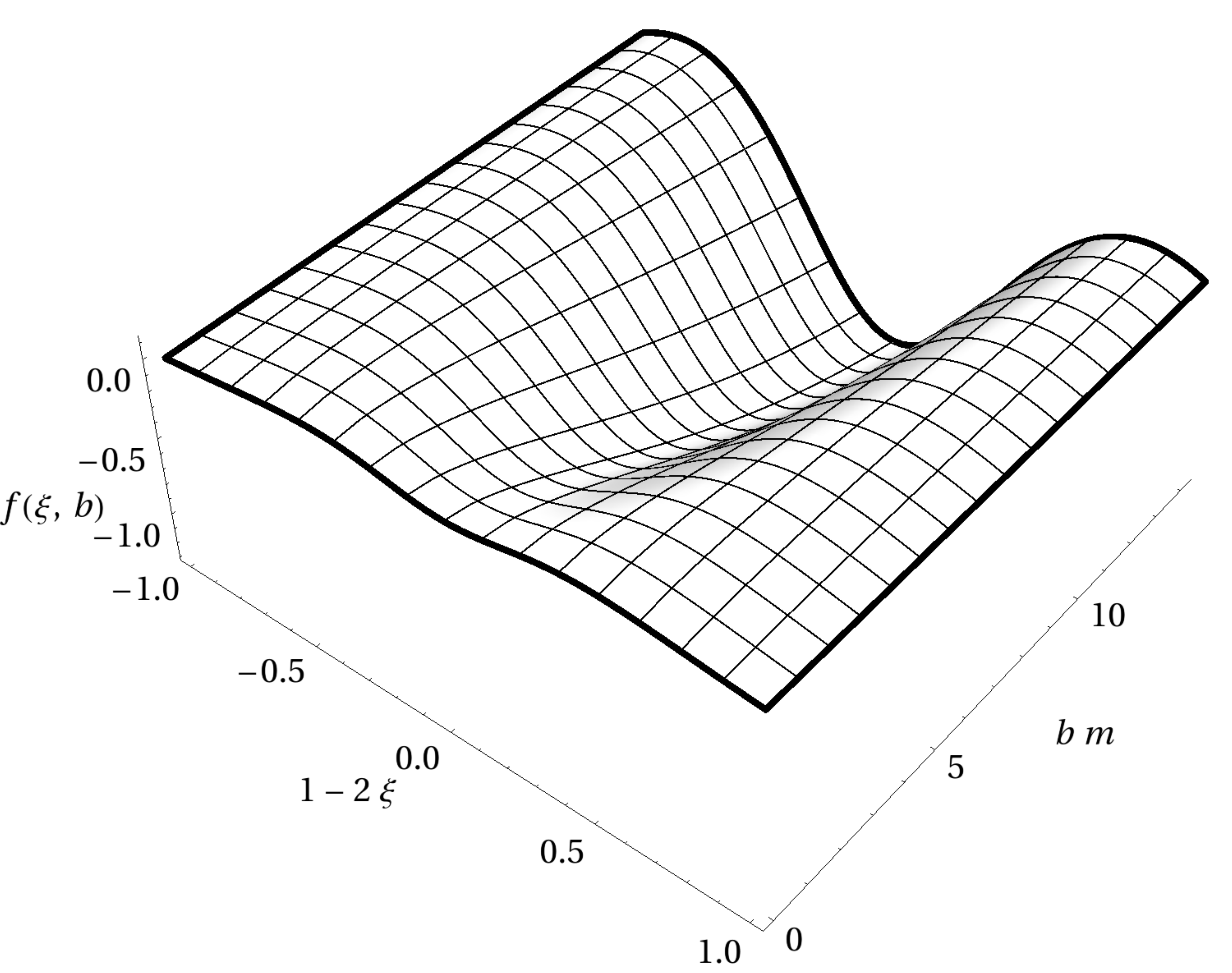}
\end{center}
\caption{
The valence functions $f(\xi,b)$ in the impact parameter space. Left panel: the ground state, corresponding to 
$B(0)$=$1.9m$, $\mu$=$0.1m$ and $\alpha_{gr}$=$6.437$.
Right panel: first-excited state, corresponding to $B(1)$=$0.22m$, $\mu$=$0.1 m$ and 
$\alpha_{gr}$=6.437~.}\label{fig:impactpara2D}
\end{figure}

The  transverse charge densities have been  thoroughly discussed by Miller in Ref.~\cite{MilARNP10},
in close relation to the elastic electromagnetic form factor. Indeed, the transverse charge density 
allows one to properly generalize   the well-known non-relativstic relation between form factor 
and density to a relativistic framework. As a matter of fact, it turns out that for a composite bosonic state, 
the form factor $F(Q^2=-q^2)$ can be written as 
\begin{equation}
F(Q^2)=\int d^2{\bf b} \,\rho({\bf b})\, \e^{-{\rm i} {\bf b}\cdot {\bf q}_\perp}
 \, ,
\end{equation}
where 
(i) the momentum transfer $q^\mu$ is  evaluated in the Breit frame with $q^+=0$,  
(ii) $Q^2={\bf q}_\perp^2$, 
(iii) ${\bf b}$ belongs to  the transverse plane, called IP space \cite{BurIJMP03}, and 
(iv) $\rho({\bf b})$ is the IP density. It has to be pointed out that the IP density
is the  sum of  contributions from all   the  LF amplitudes of the  Fock expansion 
of the interacting-system state, such that 
\begin{equation}
\rho({\bf b})=\rho_{\text{val}}({\bf b})+\,\,\text{higher Fock states densities}\,\cdots
.\end{equation}
The valence term is defined through  the valence wave function in the  IP space,
$\phi(\xi,{\bf b})$, as  follows
\begin{equation} \label{rhoval}
\rho_{\text{val}}({\bf b})=\frac{1}{4\pi}
\int^1_0 \frac{d\xi}{\xi (1-\xi)^3} \,| \phi(\xi,{\bf b}/(1-\xi))|^2
\, .
\end{equation}
with normalization (cf. Eq. (\ref{pval}))
$\int d^2{\bf b}~\rho_{\text{val}}({\bf b})=P_{val}$. In Eq. (\ref{rhoval}), the IP-space valence wave 
function is the 2D Fourier transform of $\psi(\xi,{\bf k_\perp})$, given by
\begin{equation} \label{phibgm}
\phi(\xi,{\bf b})=
\int\frac{ d^2{\bf k_\perp}}{(2\pi)^2}\; \psi(\xi,{\bf k_\perp})
{\rm e}^{{\rm i}{\bf k_
\perp}\cdot {\bf b}} 
,\end{equation}
where {$\phi(\xi,{\bf b})$ results to be symmetric with respect to $1-2\xi$, as a consequence of the 
already discussed symmetry of $g(\gamma,z;\kappa^2)$ under the transformation $z\to -z$. 
Moreover, one can deduce the general behavior for large transverse separations, $b=|{\bf b}|$, as 
illustrated in what follows. By performing the 2D Fourier transformation, the IP-space valence wave 
function can be written within NIR for the $s-$wave state as follows
\begin{equation}
\phi(\xi,b)= \frac{\xi(1-\xi)}{4\pi\sqrt{2}}~F(\xi,b)
,\label{valbt}
\end{equation}
where  
\begin{equation}\label{f(b,z)}
F(\xi,b)=\int^\infty_0 d\gamma~J_0(b \sqrt{\gamma})
\int_0^{\infty}d\gamma'~{g(\gamma',1-2\xi;\kappa^2)
\over [\gamma+\gamma' +\kappa^2+(1/2-\xi)^2M^2]^2}\, ,
\end{equation}
with $J_n(x)$ the integer-order Bessel function for $n=0$. 
Also the integration over
$\gamma$ can be  analytically carried out, leading to
\begin{equation}
F(\xi,b)
= b\,\int_0^{\infty}d\gamma~g(\gamma,1-2\xi;\kappa^2) 
\frac{K_1\left(b\,\sqrt{\gamma +\kappa^2+(1/2-\xi)^2M^2}\right)}
{\sqrt{\gamma +\kappa^2+(1/2-\xi)^2M^2}},
\label{f1(b,z)}
\end{equation}
where $K_1(x)$ is the modified Bessel function of the second kind.
The function $F(\xi,b)$ exponentially drops out for $b \to \infty$. 
Such a behavior can be
understood by a close analysis of Eq. (\ref{f1(b,z)}).}
First, from the physically-motivated request~\cite{MilARNP10} that $\phi(\xi,b)$ is 
finite for $ b\to 0$ (see also \cite{FSV2}), such that
 \begin{equation}
\phi(\xi,b=0)= 
  {\xi~(1-\xi)\over 4\pi \sqrt{2}}\int_0^{\infty}d\gamma~\frac{
g(\gamma,1-2\xi;\kappa^2)}
{\gamma+\kappa^2+(1/2-\xi)^2M^2} \, < \infty~~, 
\label{valbt1}
\end{equation}
one can deduce that $g(\gamma,1-2\xi;\kappa^2)$ must vanish for 
$\gamma \to \infty$. Therefore, the relevant interval of $\gamma$ in the
integral (\ref{f1(b,z)}) can be taken   effectively finite.  
Exploiting such an observation,
one can extract the driving exponential fall-off of $F(\xi,b)$  
in the asymptotic limit $b \rightarrow \infty$. 
In this limit $K_{1}(x)$ reads:
\begin{equation}\label{k1asymp}
K_{1}(x)|_{x\to\infty}\to\left( \frac{\pi}{2\,x} \right)^{\frac{1}{2}} \e^{-x} \, .
\end{equation}
The leading exponential behavior  in the integral (\ref{f1(b,z)}) 
comes from values of $\e^{-b\,\sqrt{\gamma+\kappa^2+(\xi-1/2)^2M^2 }}$ 
[as seen from Eq.~\ref{k1asymp})] with 
$\gamma$ close to $0$. Therefore, 
\begin{equation}\label{f2(b,z)}
F(\xi,b)|_{b\to\infty}\to~\e^{-b\,\sqrt{\kappa^2+(\xi-1/2)^2M^2}}\,f(\xi,b) ,
\end{equation}
where the exponential fall-off is singled out and the  reduced function 
$f(\xi,b)$ should decrease more smoothly 
for large values of $b$. It has to be pointed out that an  exponential fall-off 
is expected for  bound states, since it is generated by 
short range interactions, in analogy with the behavior found for the 
non-relativistic two-dimensional case. 
The above feature has been investigated by actually calculating $F(\xi,b)$, and in 
turn $f(\xi,b)$, for both ground and first-excited states. In 
Fig.~\ref{fig:impactpara2D},
$f(\xi,b)$ is presented  for  the ground (left) and 
first-excited (right) states. In both cases, we have $\mu/m=0.1$ 
and $\alpha_{gr}$=6.437. 
It is worth noting in the right panel the nice  
node structure of the excited state in the whole $\{1-2\xi,b\}$ plane.
The tail of the function $F(\xi,b)$ with respect to $b$  is studied in more 
detail through  $f(\xi,b)$,  putting in evidence that the steep fall-off  of 
the valence wave function, which  is largely taken into account  
 by the leading exponential term, included in the definition  (\ref{f2(b,z)}). This suggests
at most  a polynomial behavior in $f(\xi,b)$, which is clearly seen in the figure for $b <15/m$. 
\section{Conclusions and Perspectives}
\label{END}
We have investigated  both spectrum and excited states of the scalar Bethe-Salpeter equation, in ladder
approximation, by getting, for the first time, solutions directly in Minkowski space, within the
Nakanishi integral representation of the BS amplitude. A basic ingredient of our approach is the
exact projection of the BSE onto the null-plane (see, e.g. \cite{SalPRC00,FreFBS00}), that allows one to 
master in a simple and very effective way the singularities typical of the BS formalism.
We have considered an $s$-wave interacting system composed by two massive scalars and interacting
through a  massive scalar, extending the study of the ground state performed in Ref. \cite{FSV2} (see Ref.
\cite{carbonell1} for an analogous study within the explicitly-covariant LF approach), and
carefully analyzing the valence wave function both in Minkowski and impact-parameter spaces.

Within the numerical accuracy of our approach, we have found a finite 
number of excited states for non-zero exchanged-scalar mass, and we have successfully 
compared our results with the corresponding ones 
obtained in the Euclidean space, where obviously  NIR is not assumed. A detailed  study of the
valence wave function structure has been carried out in the plane $(\xi,k_\perp)$, showing the 
expected node structure of the first and second excited states. Furthermore,
our investigation, both analytic and quantitative, of the transverse-momentum amplitude has allowed
 to remarkably show the
equivalence of the quantity evaluated both in Euclidean and Minkowski spaces. This  further strengthens 
the  reliability of the approach based on NIR for solving the BSE,  already applied to
fermionic systems \cite{carbonell3}, kernels beyond the ladder one \cite{carbonell2} and in the
zero-energy scattering case \cite{FSV3}. 
Finally, we also explored the asymptotic properties of the impact-parameter space valence wave function 
for large transverse distances, where an exponential fall-off  was singled out 
(similar to the non-relativistic case in the 3D Euclidean space) and quantitatively tested for 
the excited states. 

In perspective, the present study encourages the extension of  the approach based on the NIR to excited 
states of actual physical systems, as well as to explore results obtained for other dynamical quantities within a
wider and deeper comparison between Minkowski and Euclidean calculations.

{\it Acknowledgments.} 
We thank the Brazilian agencies Coordena\c c\~ao de Aperfei\c coamento de Pessoal de N\'ivel Superior
(CAPES), Conselho Nacional de Desenvolvimento Cient\'ifico e Tecnol\'ogico (CNPq) and
Funda\c c\~ao de Amparo \`a Pesquisa do Estado de S\~ao Paulo (FAPESP) for partial support.
MV and GS acknowledge the warm hospitality of
the Instituto Tecnol\'ogico de Aeron\'autica, S\~ao Jos\'e dos Campos where part of this work was performed.

\end{document}